\documentclass[
    reprint,
    amsmath,amssymb,
]{revtex4-2}

\usepackage{graphicx}
\usepackage{dcolumn}
\usepackage[dvipsnames]{xcolor}
\usepackage{bm}
\usepackage{hyperref}
\usepackage{url}
\usepackage{upgreek}
\usepackage{gensymb}

\hypersetup{
    colorlinks=true,
    linkcolor=blue,
    filecolor=magenta,      
    urlcolor=cyan,
    pdftitle={Overleaf Example},
    pdfpagemode=FullScreen,
    }
    
\begin{document}
\preprint{APS/123-QED}

\title{Single-Shot Flow Spectroscopy of a Polariton Condensate: Kibble--Zurek and Kolmogorov-Like Scaling}

\author{Ivan Krasionov$^{1}$}
\author{Anton Putintsev$^{1}$}
\author{Maksim Kolker$^{1}$}
\author{Tamsin Cookson$^{1}$}
\author{Sergey Alyatkin$^{1}$}
\author{Pavlos G. Lagoudakis$^{1}$}

\thanks{Contact author: p.lagoudakis@skoltech.ru}

\affiliation{$^1$Hybrid Photonics Laboratory, Skolkovo Institute of Science and Technology, Territory of Innovation Center Skolkovo, Bolshoy Boulevard 30, building 1, 121205 Moscow, Russia}

\date{\today}

\begin{abstract}
Quantized vortices are fundamental topological excitations of quantum fluids. We report single-shot interferometric measurements of spontaneous vortex nucleation in a room-temperature organic exciton--polariton condensate. From hundreds of independent realizations we find random vortex-core positions and unbiased circulation, consistent with intrinsically stochastic, unpinned defect formation. The mean vortex number scales with pump power above threshold with an exponent consistent with Kibble--Zurek freeze-out in a driven--dissipative condensate. Using reconstructed phase maps we obtain single-shot flow fields, compute the incompressible component, and extract kinetic-energy spectra. Vortex-containing realizations develop a robust Kolmogorov-like segment with $E_{\rm inc}(k)\propto k^{-5/3}$ over a finite $k$ range, indicating the onset of turbulent spectral scaling in a quantum fluid of light. These results establish single-shot access to phase and flow as a direct route to quantifying stochastic defect formation and emerging turbulence in polariton condensates.
\end{abstract}
                              
\maketitle

\section*{Introduction}
\vspace*{-0.25cm}
Quantized vortices are topological defects of a macroscopic wavefunction and a central dynamical ingredient of superfluids and Bose--Einstein condensates \cite{vinen2002quantum,barenghi2001quantized,weiler2008spontaneous}. In nonequilibrium condensates, vortices can nucleate spontaneously during the phase transition, providing a direct window into symmetry breaking, phase ordering, and turbulent relaxation.

Exciton--polaritons---hybrid light--matter quasiparticles in microcavities---form driven--dissipative condensates with direct optical access to density and phase \cite{carusotto2013quantum,deng2002condensation}. In organic microcavities, condensation and nonlinear hydrodynamics can be accessed at room temperature under ultrafast optical pumping \cite{cookson2017yellow,peng2022room,christopoulos2007room}. Vortices can be imprinted by structured excitation \cite{boulier2015vortex} or arise spontaneously from disorder, flows, or the condensation process itself \cite{lagoudakis2008quantized,panico2023onset,zamora2020kibble}.

When a continuous phase transition is driven at a finite rate, non-equilibrium dynamics can lead to the formation of topological defects as described by the Kibble--Zurek mechanism (KZM) \cite{del2014universality}. Close to the critical point, the correlation length $\xi$ and relaxation time $\tau_{\rm rel}$ scale as $\xi(\epsilon)\sim|\epsilon|^{-\nu}$ and $\tau_{\rm rel}(\epsilon)\sim|\epsilon|^{-z\nu}$, where $\epsilon$ is the reduced distance to criticality, $\nu$ is the correlation-length exponent, and $z$ is the dynamic critical exponent. The growth of $\tau_{\rm rel}$ implies a loss of adiabaticity during the drive, fixing a finite freeze-out correlation length $\hat{\xi}$ and hence a characteristic defect spacing. Consequently, the vortex density obeys
\begin{equation}
  n_V \sim \hat{\xi}^{-(D-d)} \sim \tau_Q^{-(D-d)\nu/(z\nu+1)},
\end{equation}
where $D$ is the spatial dimension and $d$ is the defect dimension ($d=0$ for vortices in 2D), and $\tau_Q$ is the characteristic quench time \cite{huang2014kibble}. For the present driven--dissipative condensate, we use the pump power relative to threshold as an experimentally accessible proxy for the effective quench rate when testing KZM-type scaling.

In two-dimensional (2D) systems, the incompressible kinetic energy of a turbulent quantum flow is expected to cascade from small to large length scales with a Kolmogorov power-law spectrum $E(k)\propto k^{-5/3}$ \cite{kraichnan1967inertial, bradley2012energy}. Evidence of vortex clustering and $-5/3$ scaling has been reported in ultracold atomic BECs \cite{skrbek2012developed, kobayashi2008quantum, parker2004controlled, neely2010observation, zhao2025kolmogorov} and in polariton condensates \cite{berloff2010turbulence, ferrini2025driven, panico2023onset, koniakhin20202d, comaron2025dynamics}.

\begin{figure*}[t]
  \centering
  \includegraphics[width=\textwidth]{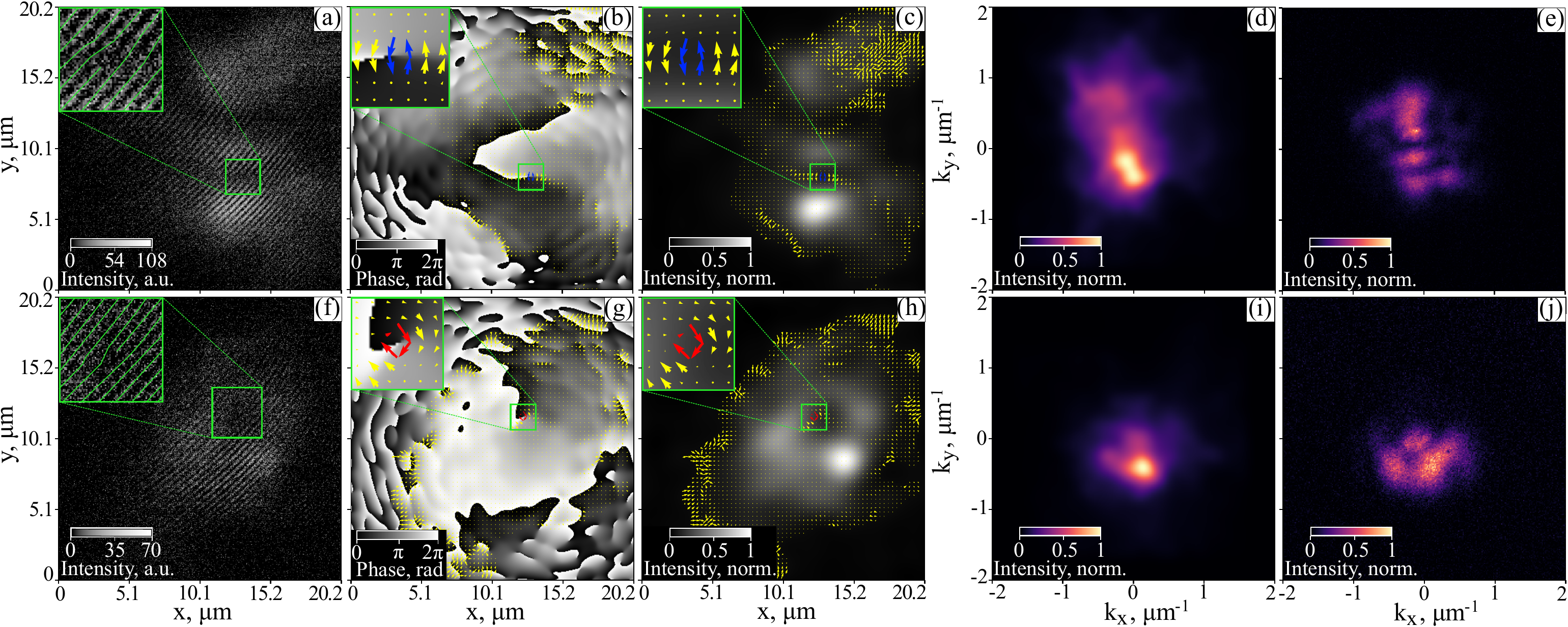}
  \caption{\label{fig:1}
  \textbf{Single-shot interferometry and flow reconstruction.} Two independent single-shot measurements at the same sample location, shown in panels (a)–(e) and (f)–(j), respectively. \textbf{(a),(f)} Single-shot interferograms, each containing a spontaneously formed vortex (fork-like dislocation in the interference fringes, inset). Green curves trace the fringes as a guide to the eye. \textbf{(b),(g)} Reconstructed real-space phase profiles obtained via an off-axis digital holography algorithm. Yellow arrows indicate the local wavevector field $\mathbf{k}(\mathbf{r},t)=\nabla\phi(\mathbf{r},t)$ in regions where the condensate intensity exceeds 5\% of its peak value. Blue and red arrows denote vortex cores with counterclockwise and clockwise circulation, respectively; arrow lengths are proportional to $|\mathbf{k}|$. \textbf{(c),(h)} Reconstructed real-space condensate density profiles. \textbf{(d),(i)} Far-field (momentum-space) intensity distributions reconstructed from (c) and (h), respectively. \textbf{(e),(j)} Experimentally measured far-field intensity images.}
\end{figure*}

Previous studies of vortex formation in polariton condensates have predominantly relied on time-integrated photoluminescence, which averages over many stochastic realizations and can obscure nucleation statistics and transient dynamics \cite{lagoudakis2008quantized}. Single-shot interferometry avoids ensemble averaging, allowing each realization to be reconstructed independently and enabling statistically meaningful tests of spontaneous defect formation and emergent flow patterns.

Here we use single-shot pulsed excitation and phase-resolved interferometric imaging to track spontaneous vortex nucleation in a room-temperature organic polariton condensate. We show that vortex positions and circulation are unbiased, consistent with unpinned nucleation. By varying excitation above threshold, we test Kibble--Zurek scaling of the mean vortex number. Finally, from single-shot flow fields we extract incompressible kinetic-energy spectra and identify a Kolmogorov-like $k^{-5/3}$ segment in vortex-containing realizations.

\vspace*{-0.25cm}
\section*{Results and Discussion}
\vspace*{-0.25cm}

A critical challenge in single-shot photoluminescence measurements is the low emission intensity of a single condensate realization, which ordinarily would necessitate integrating the signal over many realizations. We use a BODIPY-Br organic microcavity (fabrication details in Supplemental Materials (SM) Sec.~\ref{sec:I}) that supports room-temperature condensation and excite it with $\sim$250\,fs pulses at 400\,nm focused to a $\sim$23~$\upmu$m (FWHM) spot. 
The condensate emission is split into two detection paths. In the first, we record a single-shot interferogram using a Mach--Zehnder interferometer, where one arm expands and inverts the condensate image about its center and serves as a reference wave. Representative single-shot interferometric patterns acquired at the same sample location are presented in Figs.~\ref{fig:1}(a)\&(f). 

We apply an off-axis digital holography algorithm to reconstruct both the condensate phase $\phi(\textbf{r},t)$ \mbox{[Figs.~\ref{fig:1}(b)\&(g)]} and its intensity distribution \mbox{[Figs.~\ref{fig:1}(c)\&(h)]} (SM Sec.~\ref{sec:II}). The local wavevectors $\textbf{k}(\textbf{r},t) = \nabla \phi (\textbf{r},t)$, depicted by yellow arrows, are computed from the phase maps (thresholded at 5\% of peak intensity). We obtain an intensity-weighted momentum distribution \mbox{[Figs.~\ref{fig:1}(d)\&(i)]} that agrees with the simultaneously recorded far-field images \mbox{[Figs.~\ref{fig:1}(e)\&(j)]}, validating the single-shot reconstruction.

\begin{figure}[t]
  \centering
  \includegraphics[scale=0.38]{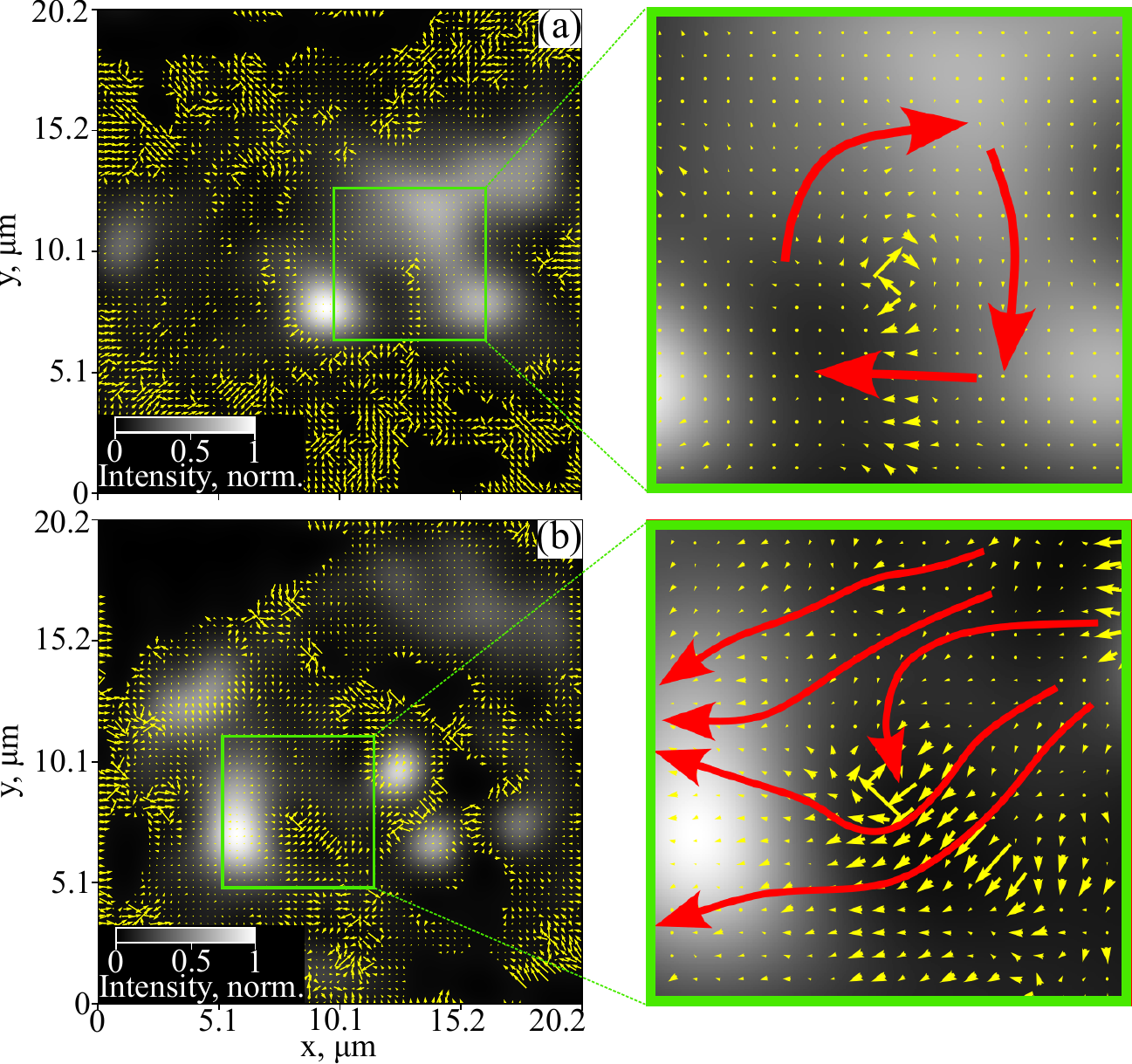}
  \caption{\label{fig:2}
  \textbf{Flow maps.} Representative single-shot phase-retrieval results from different sample locations. \textbf{(a)} Vortex core at the interface between two large-scale counter-propagating (or locally rotating) flow domains. \textbf{(b)} Vortex within an approximately co-directional flow. Yellow arrows: reconstructed wavevector (flow) field. Red arrows (insets): overall flow directions of the neighboring domains.}
\end{figure}

Single-shot phase retrieval reveals not only the local $2\pi$ winding at a vortex core but also the global phase and flow configuration of each realization. Vortex cores exhibit a strong spatial correlation with extended interfaces separating regions of oppositely directed superflow \mbox{[Fig.~\ref{fig:2}(a)]}: in $\approx98\%$ of all identified vortices (203 out of 207), the core lies on or adjacent to a boundary between two large-scale flow domains, where an interface is identified by a sign change in the local coarse-grained flow direction. Only $\approx2\%$ (4 out of 207) appear within approximately co-directional flow regions \mbox{[Fig.~\ref{fig:2}(b)]}, consistent with rare events influenced by weak scattering from local inhomogeneities.

\begin{figure}[t]
  \centering
  \includegraphics[scale=0.38]{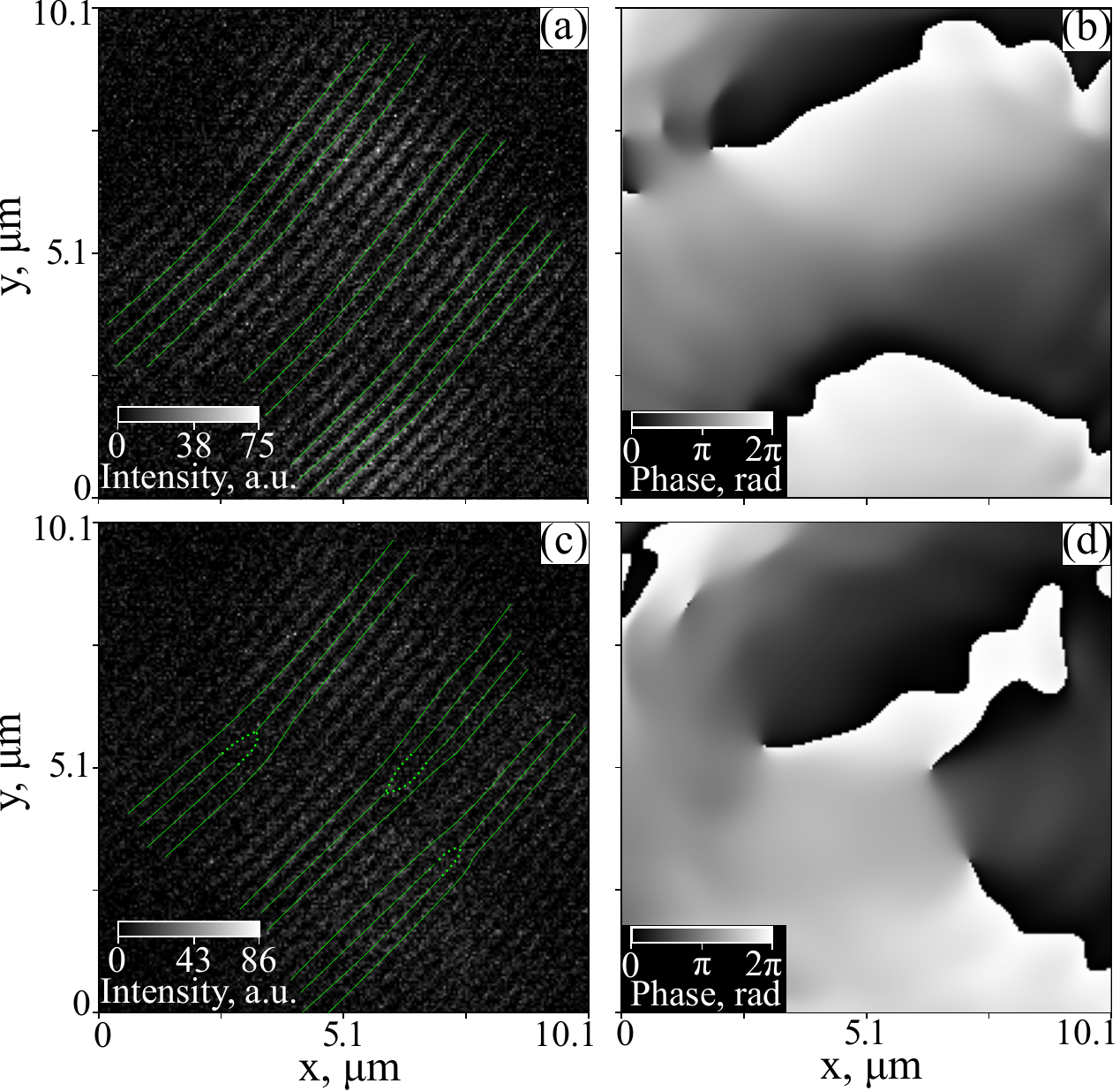}
  \caption{\label{fig:3}
  \textbf{Vortex (un)pinning at a fixed location.} Two single-shot interferograms taken in succession at the same sample spot. \textbf{(a)} Interference fringes showing no vortex (uniform fringe pattern) at the areas highlighted by green curves. \textbf{(c)} Fringes from the immediately following shot, in which a vortex appears (fork-like dislocation). \textbf{(b),(d)} Reconstructed phase maps corresponding to (a) and (c), respectively.}
\end{figure}

The stochastic character of vortex formation is further evidenced by the absence of reproducible nucleation sites. Across 458 single-shot measurements acquired at 17 distinct sample positions, the vortex number varies from 0 to 3 per realization, and vortex positions fluctuate from shot to shot, showing no statistically reproducible nucleation hotspots beyond the pump-envelope weighting (SM Sec.~\ref{sec:III}). Across the full dataset we identify 98 vortices and 109 antivortices, consistent with unbiased circulation within counting uncertainty. Using a conservative coincidence criterion based on the measured sample vibration amplitude $(\approx0.37~\upmu\mathrm{m})$, we find no indication of systematic pinning: vortices can appear or disappear at the same nominal location in consecutive shots, and their circulation can occasionally reverse, see Fig.~\ref{fig:3}. To reduce the influence of any single static defect, we translated the excitation spot to a new location after a limited number of shots at each position (typically 20--30), which also helped to mitigate local photo-degradation of the organic film.

To quantify nucleation, we analyze the mean vortex number $n_V$ versus effective quench rate controlled by the pump power. At fixed pulse duration and spot size, increasing $P/P_{\rm thr}$ accelerates the build-up of macroscopic coherence and thus provides a monotonic experimental proxy for the effective quench rate \cite{Matuszewski2014, Chen2022}. The threshold is obtained from the input--output nonlinearity and linewidth collapse (SM Sec.~\ref{sec:IV}); for the $\sim$23~$\upmu$m spot we find $P_{\rm thr}=551~\upmu\mathrm{J\,cm^{-2}}$. In a mean-field benchmark for a parabolic dispersion and a linear-in-energy relaxation rate \cite{solnyshkov2016kibble, solnyshkov2021kibble}, one expects $\nu=1/2$ and $z=2$, yielding the Kibble--Zurek exponent $\alpha=(D-d)\nu/(z\nu+1)=1/2$ for point defects in $D=2$.

\begin{figure}[t]
  \centering
  \includegraphics[scale=0.5]{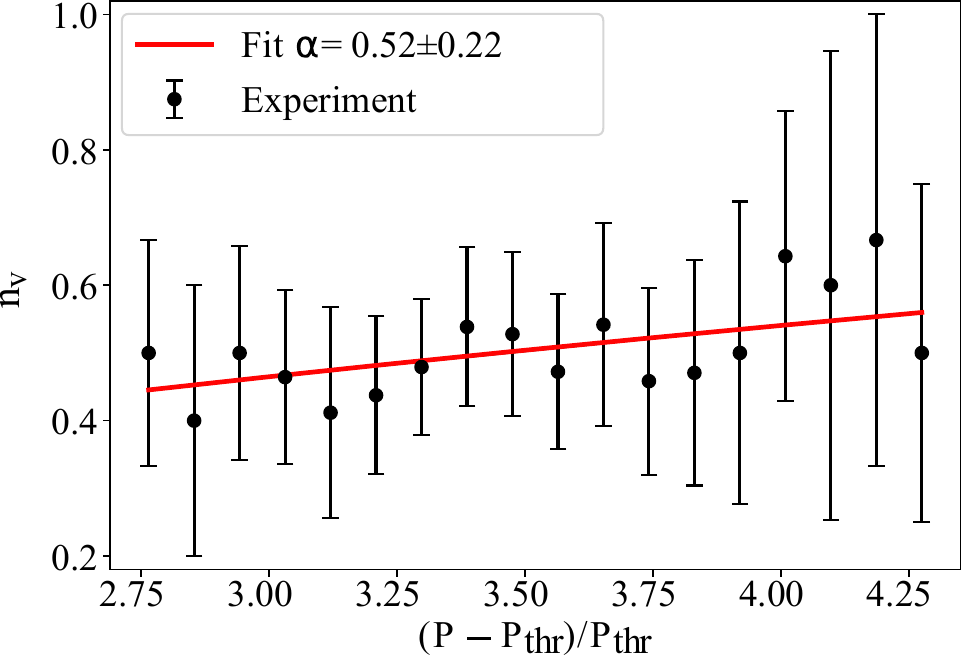}
  \caption{\label{fig:4}
  \textbf{Kibble--Zurek scaling of vortex number.} Average vortex count $n_V$ versus relative pump power. Error bars indicate uncertainty from Poisson statistics. Red line: power-law fit over $2.75 \le (P-P_{\mathrm{thr}})/P_{\mathrm{thr}} \le 4.3$ (SM Fig.~\ref{fig:S3}), yielding $\alpha\simeq0.5$ as expected for 2D point defects in the mean-field benchmark.}
\end{figure}

The experimental results are shown in Fig.~\ref{fig:4}. For each pump power we estimate $n_V$ and its uncertainty assuming Poisson counting statistics, and fit the result by a power law over the range $2.75 \le (P-P_{\mathrm{thr}})/P_{\mathrm{thr}} \le 4.3$ (SM Fig.~\ref{fig:S3}). For the finest binning (with at least two data points per bin), the fit yields $\alpha=0.52\pm0.22$. Repeating the fit over 30 binning realizations gives $\alpha=0.50\pm0.25$ (SM Fig.~\ref{fig:S4}), capturing counting noise and binning sensitivity. Above $(P-P_{\mathrm{thr}})/P_{\mathrm{thr}}=4.3$ we observe a decrease in vortex number, consistent with saturation effects (SM Sec.~\ref{sec:IV}). Related KZM studies in polaritons with additional Rashba--Dresselhaus spin--orbit coupling report modified scaling exponents \cite{solnyshkov2025kibble}.

\begin{figure}[t]
  \centering
  \includegraphics[scale=0.5]{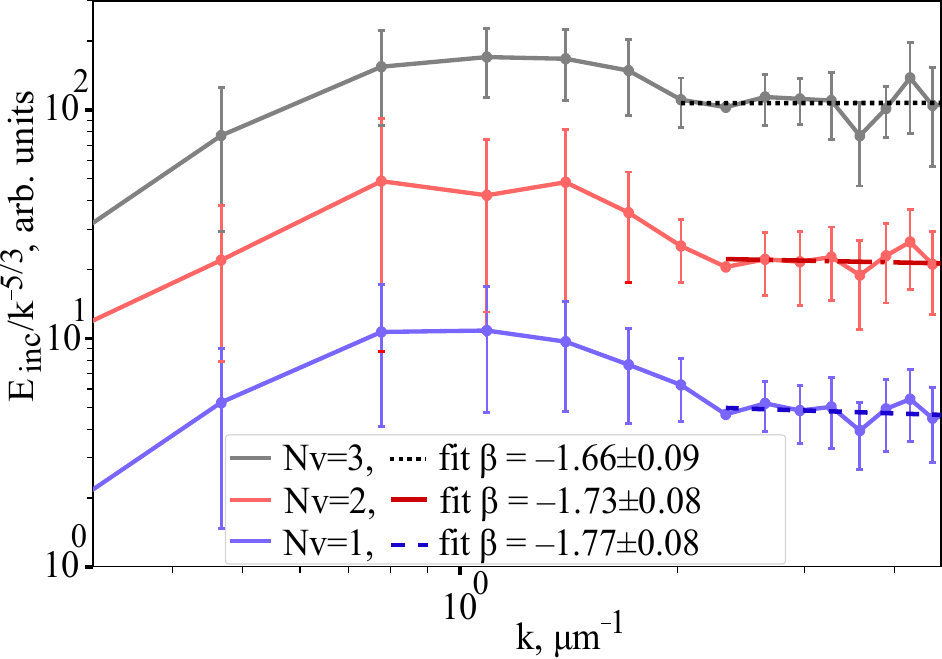}
  \caption{\label{fig:5}
  \textbf{Incompressible kinetic-energy spectra.} Log--log plots of the angle-averaged incompressible spectrum $E_{\rm inc}(k)$ for single-shot realizations containing $N_V=1$ (blue), 2 (green), or 3 (red) vortices. Curves show group-averaged spectra and are vertically offset for clarity. Dashed lines are power-law fits performed over an intermediate-$k$ interval between the system-size scale ($k\sim2\pi/L$) and the optical-resolution scale ($k\sim\pi/\delta r$), showing an approximate $k^{-5/3}$ dependence.}
\end{figure}

Finally, we analyze the flow-field spectra in vortex-containing shots. For each realization we compute the velocity field, project onto its divergence-free component by a Helmholtz decomposition in $k$ space, and azimuthally average to obtain $E_{\rm inc}(k)$ \cite{koniakhin20202d}. Grouping shots by vortex number $N_V=1$--3, we find that $E_{\rm inc}(k)$ develops an extended segment with an effective slope close to $-5/3$, see Fig.~\ref{fig:5}. Local fits yield $\beta=-1.77\pm0.08$ ($N_V=1$), $-1.73\pm0.09$ ($N_V=2$), and $-1.66\pm0.09$ ($N_V=3$). While the finite system size and $N_V\le3$ preclude establishing a flux-supported inertial range, the appearance of a robust Kolmogorov-like slope already in the few-vortex regime indicates the onset of turbulent spectral scaling in a quantum fluid of light.

In conclusion, we demonstrate single-shot interferometric imaging of spontaneous vortex nucleation in a room-temperature organic polariton condensate. Single-shot phase reconstruction reveals both vortex cores and the global flow configuration, with cores predominantly appearing at interfaces between large-scale counter-flow domains. Vortex positions and circulation are unbiased, consistent with stochastic, unpinned nucleation. The mean vortex number scales with pump power above threshold with a Kibble--Zurek exponent $\alpha\simeq 1/2$. From single-shot flow fields we extract incompressible spectra showing a robust $k^{-5/3}$-like segment in few-vortex realizations; although finite size and $N_V\le 3$ preclude establishing a flux-supported inertial range, the observed scaling provides a starting point for mapping the emergence of turbulence with increasing vortex number and clustering.

\bibliographystyle{apsrev4-2}
\bibliography{Bibliography}

\providecommand{\noopsort}[1]{}
\providecommand{\singleletter}[1]{#1}

\setcounter{equation}{0}
\setcounter{figure}{0}
\setcounter{section}{0}
\setcounter{subsection}{0}
\newcolumntype{P}[1]{>{\centering\arraybackslash}p{#1}}
\newcolumntype{M}[1]{>{\centering\arraybackslash}m{#1}}
\renewcommand{\theequation}{S\arabic{equation}}
\renewcommand{\thefigure}{S\arabic{figure}}
\renewcommand{\baselinestretch}{1}
\onecolumngrid
\newpage
\vspace{1cm}
\begin{center}
\Large \textbf{Supplemental Materials}
\end{center}
\section{Sample preparation}
\label{sec:I}

A polymer matrix solution was prepared using polystyrene (PS) with an average molecular weight of $\approx 192\,000$ in toluene at a concentration of $35~\rm mg\,mL^{-1}$. The PS/toluene solution was heated to $70\,\degree\mathrm{C}$ and stirred for 30\,min. BODIPY-Br was then added at a concentration of $10\%$ by mass. Noncavity films for absorption, photoluminescence, and ASE measurements were spin-cast onto quartz-coated glass substrates.

For microcavity fabrication, a bottom 10-pair DBR of $\rm SiO_2/Nb_2O_5$ was deposited onto quartz-coated glass substrates using ion-assisted electron-beam evaporation ($\rm Nb_2O_5$) and thermal evaporation ($\rm SiO_2$). A 186\,nm-thick BODIPY-Br active layer was spin-coated on top of the bottom mirror. A second 8-pair DBR was deposited on top of the organic layer, with the ion gun kept off during the first few layers to avoid damage to the organic film.

\section{Single-shot phase and amplitude reconstruction}
\label{sec:II}

In this section we describe reconstruction of the condensate phase and amplitude from single-shot interference patterns.

\begin{figure}[h]
  \centering
  \includegraphics[scale=0.45]{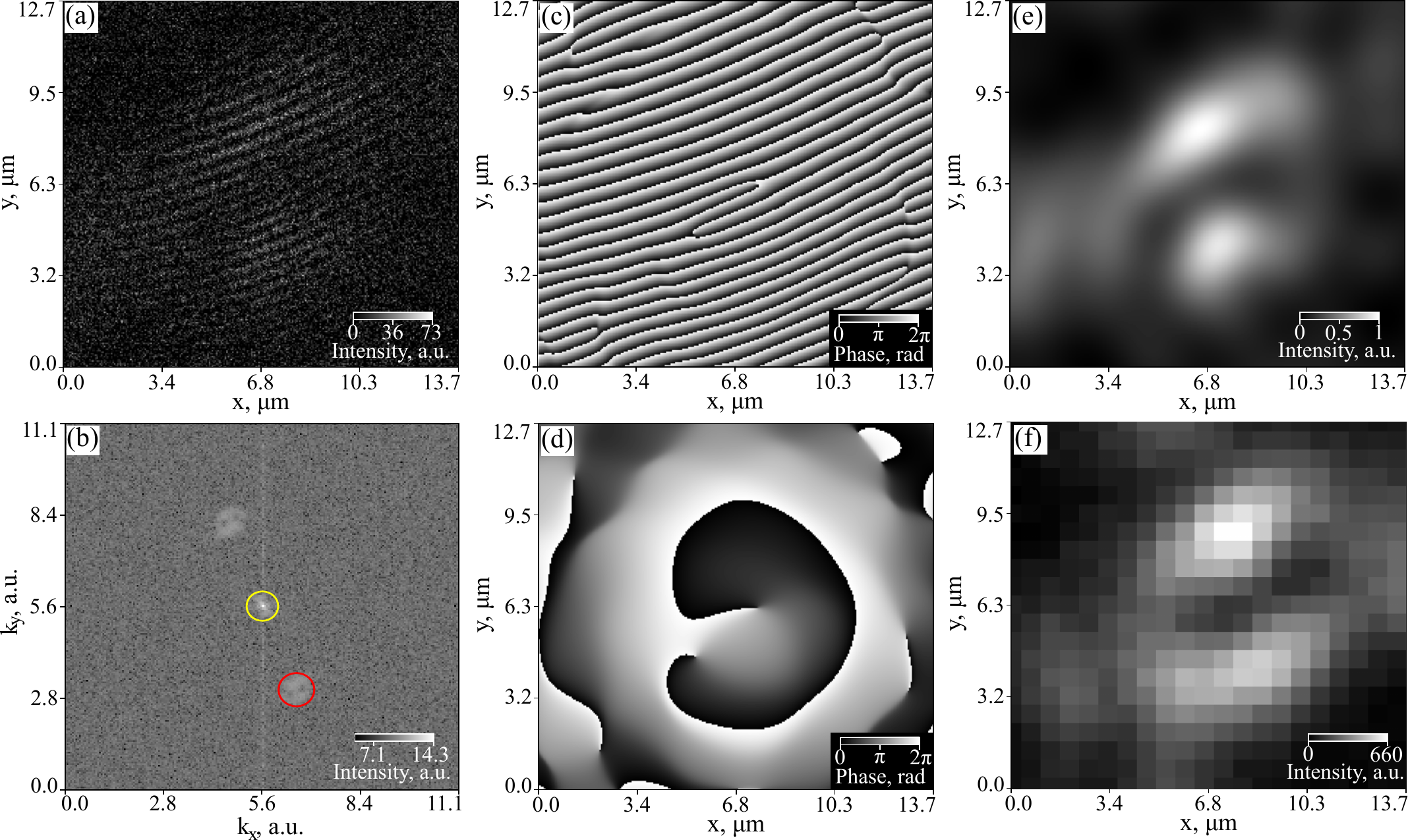}
  \caption{\label{fig:S1}
  \textbf{Single-shot phase and amplitude reconstruction by off-axis digital holography.}
  \textbf{(a)} Representative real-space interference pattern.
  \textbf{(b)} Magnitude of the 2D Fourier transform; the central component $\tilde{a}(k_x,k_y)$ and the two complex-conjugate sidebands $\tilde{c}$ and $\tilde{c}^\ast$ are separated by the off-axis carrier (circle indicates the selected sideband).
  \textbf{(c)} Phase of the inverse Fourier transform of the isolated sideband, $c'(x,y)=2\pi(k_x^c x+k_y^c y)-\phi(x,y)$.
  \textbf{(d)} Reconstructed condensate phase $\phi(x,y)$ after subtraction of the reference phase.
  \textbf{(e)} Fringe-suppressed intensity obtained by retaining only the central Fourier component, $a(x,y)=A^2(x,y)+B^2(x,y)$ (condensate intensity up to an additive offset).
  \textbf{(f)} Independently measured real-space intensity.}
\end{figure}

We employ a Mach--Zehnder interferometer in which one arm produces an expanded and center-inverted image of the condensate that serves as a reference field after recombination. A representative interference pattern is shown in Fig.~\ref{fig:S1}(a). Using off-axis digital holography we reconstruct the complex condensate field
\begin{equation}
  \psi(x,y) = A(x,y)\exp[i\phi(x,y)],
\end{equation}
thereby obtaining the phase $\phi(x,y)$ and the real-space intensity $I(x,y)=|A(x,y)|^2$.

Locally, the expanded reference beam can be approximated as an oblique plane wave,
\begin{equation}
  \psi_{\rm ref}(x,y) = B(x,y)\exp\!\left[i2\pi(k_x^c x+k_y^c y)\right],
\end{equation}
where $\mathbf{k}_c=(k_x^c,k_y^c)$ is the spatial frequency of the interference fringes set by the off-axis tilt. The recorded interference pattern is the intensity of the superposition,
\begin{equation}
  I(x,y)=|\psi(x,y)+\psi_{\rm ref}(x,y)|^2=a(x,y)+c(x,y)+c^\ast(x,y),
\end{equation}
with
\begin{equation*}
\left\{
\begin{aligned}
&a(x,y) = A^2(x,y) + B^2(x,y),\\
&c(x,y) = A(x,y)B(x,y)\exp[i(2\pi k_x^c x+2\pi k_y^c y-\phi(x,y))],\\
&c^\ast(x,y) = A(x,y)B(x,y)\exp[-i(2\pi k_x^c x+2\pi k_y^c y-\phi(x,y))].
\end{aligned}
\right.
\end{equation*}

We compute the 2D Fourier transform $\mathcal{F}(\cdot)$ of the interference pattern [Fig.~\ref{fig:S1}(b)]:
\begin{equation}
  \tilde{I}(k_x,k_y)=\mathcal{F}[I(x,y)]=\tilde{a}(k_x,k_y)+\tilde{c}(k_x,k_y)+\tilde{c}^\ast(k_x,k_y).
\end{equation}
The off-axis carrier shifts the interference contributions away from the origin in the spatial-frequency domain. The term $\tilde{a}$ is centered at $(0,0)$ and corresponds to the sum of the individual beam intensities. The cross-terms $\tilde{c}$ and $\tilde{c}^\ast$ form two complex-conjugate sidebands centered at $(k_x^c,k_y^c)$ and $(-k_x^c,-k_y^c)$ and contain the condensate phase information.

Isolating a single sideband (e.g., $\tilde{c}(k_x,k_y)$) by spatial filtering and suppressing the remaining components, the inverse Fourier transform of the filtered spectrum yields a complex field with phase
$c'(x,y)=2\pi(k_x^c x+k_y^c y)-\phi(x,y)$ [Fig.~\ref{fig:S1}(c)], so that subtraction of the reference phase reveals the condensate phase $\phi(x,y)$ [Fig.~\ref{fig:S1}(d)].

To obtain the condensate intensity profile, we retain only the central component $\tilde{a}(k_x,k_y)$ in the Fourier domain, which suppresses the fringe modulation and removes the interference terms. The inverse Fourier transform of the filtered spectrum then yields $a(x,y)=A^2(x,y)+B^2(x,y)$ [Fig.~\ref{fig:S1}(e)]. Assuming that the reference intensity is locally uniform, $B^2(x,y)\approx{\rm const}$ over the region of interest, this procedure provides the condensate intensity distribution up to an additive offset. The reconstructed intensity profile agrees well with the independently measured real-space intensity [Fig.~\ref{fig:S1}(f)].

\newpage
\section{Supplementary vortex statistics}
\label{sec:III}

In this section we provide supplementary statistics supporting the stochastic nature of vortex nucleation.

\begin{figure}[h]
  \centering
  \includegraphics[scale=0.5]{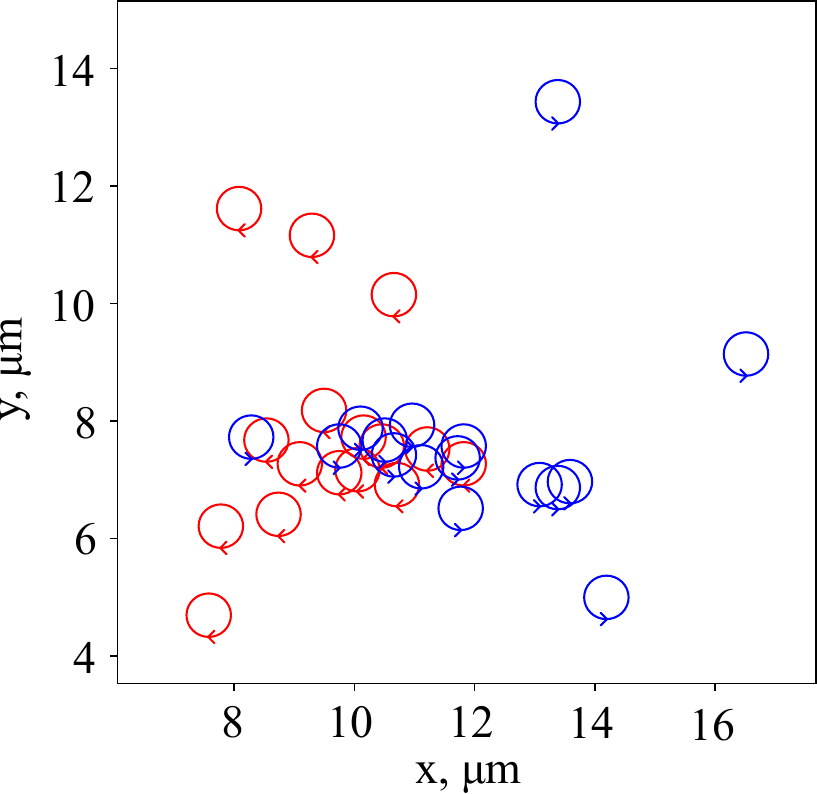}
  \caption{\label{fig:S2}
  \textbf{Vortex position distribution} for the sample location with the largest number of single-shot excitations. Red and blue markers denote vortices with clockwise and counterclockwise circulation, respectively. The radius of the circle corresponds to the pump-position jitter of $0.37~\upmu\mathrm{m}$.}
\end{figure}

Figure~\ref{fig:S2} shows the spatial distribution of vortex cores for the sample position with the largest number of single-shot excitations in our dataset. Red and blue markers denote opposite circulation (clockwise and counterclockwise, respectively). The circle radius corresponds to the measured pump-position jitter (sample vibration amplitude) of $0.37~\upmu\mathrm{m}$, which sets our conservative coincidence criterion for identifying repeated nucleation at the ``same'' nominal location. Under this criterion we observe no reproducible nucleation sites: cores vary in position from shot to shot and can change circulation sign, and many shots show no vortex despite similar excitation conditions. Across the full dataset we identify 98 vortices and 109 antivortices, consistent with unbiased circulation within counting uncertainty (a binomial fluctuation around 50:50).

\begin{figure}[b!]
  \centering
  \includegraphics[scale=0.55]{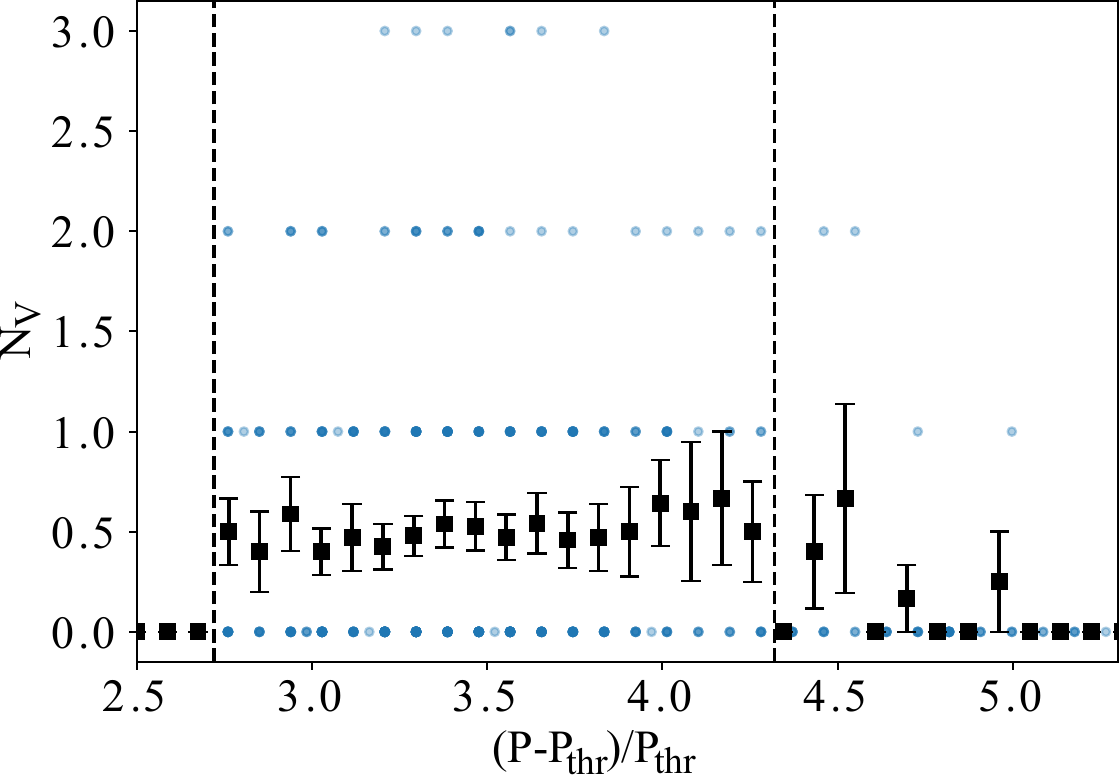}
  \caption{\label{fig:S3}
  \textbf{Vortex number distribution} as a function of pump power (blue circles). Vertical dashed lines indicate the range used for quantitative analysis. Black squares show the mean vortex number with error bars given by Poisson statistics.}
\end{figure}

Figure~\ref{fig:S3} summarizes the vortex-number statistics versus pump power. Blue markers show the vortex counts extracted from individual single-shot realizations, while black squares report the mean vortex number with Poisson error bars; the vertical dashed lines indicate the pump-power range used for quantitative analysis in the main text. In the low-to-intermediate pumping regime the mean vortex number increases monotonically, consistent with a faster build-up of coherence: stronger pumping accelerates formation dynamics, reduces the effective freeze-out length scale (smaller phase-ordering domains), and increases the probability of topological defect formation. Above $(P-P_{\rm thr})/P_{\rm thr}=4.3$ we observe a decrease in the average vortex number, consistent with saturation effects.

\begin{figure}[h!]
  \centering
  \includegraphics[scale=0.55]{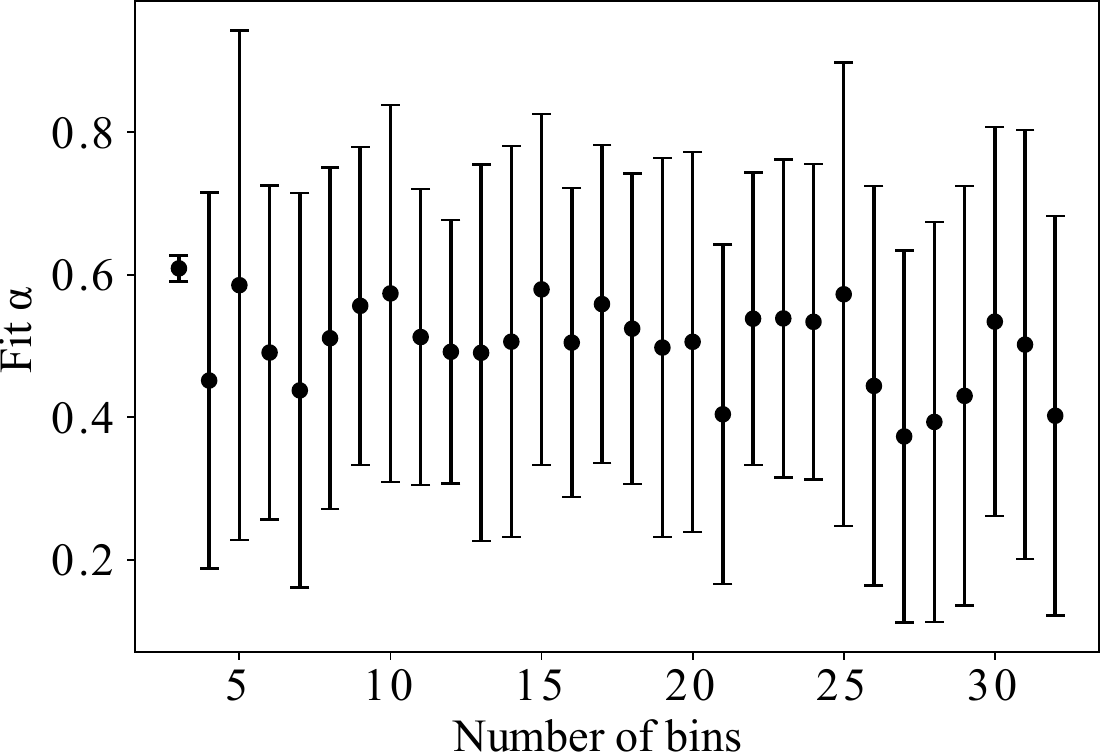}
  \caption{\label{fig:S4}
  \textbf{Binning sensitivity of the KZM exponent.} Dependence of the fitted scaling exponent $\alpha$ on the number of bins used for the power-law fit. Averaging over 30 binning realizations yields $\alpha=0.50\pm0.25$, with the uncertainty estimated as $\sigma_{\rm mean}=\sqrt{\frac{1}{N}\sum_{i=1}^{N}\sigma_i^2}$.}
\end{figure}

We assess the robustness of the extracted Kibble--Zurek exponent $\alpha$ to binning choices, see Fig.~\ref{fig:S4}. Because the vortex number is discrete (0--3 per shot) and the number of realizations per power is finite, the fitted slope can be sensitive to bin edges and occupancy. To quantify this uncertainty, we repeat the fit over 30 independent binning realizations and report the resulting distribution of $\alpha$. This yields $\alpha=0.50\pm0.25$, where the quoted uncertainty captures both counting noise and binning sensitivity.

\newpage
\newpage
\section{Threshold determination}
\label{sec:IV}

In this section we characterize the polariton condensate emission and determine the threshold pump fluence $P_{\rm thr}$.

\begin{figure}[h]
  \centering
  \includegraphics[scale=0.55]{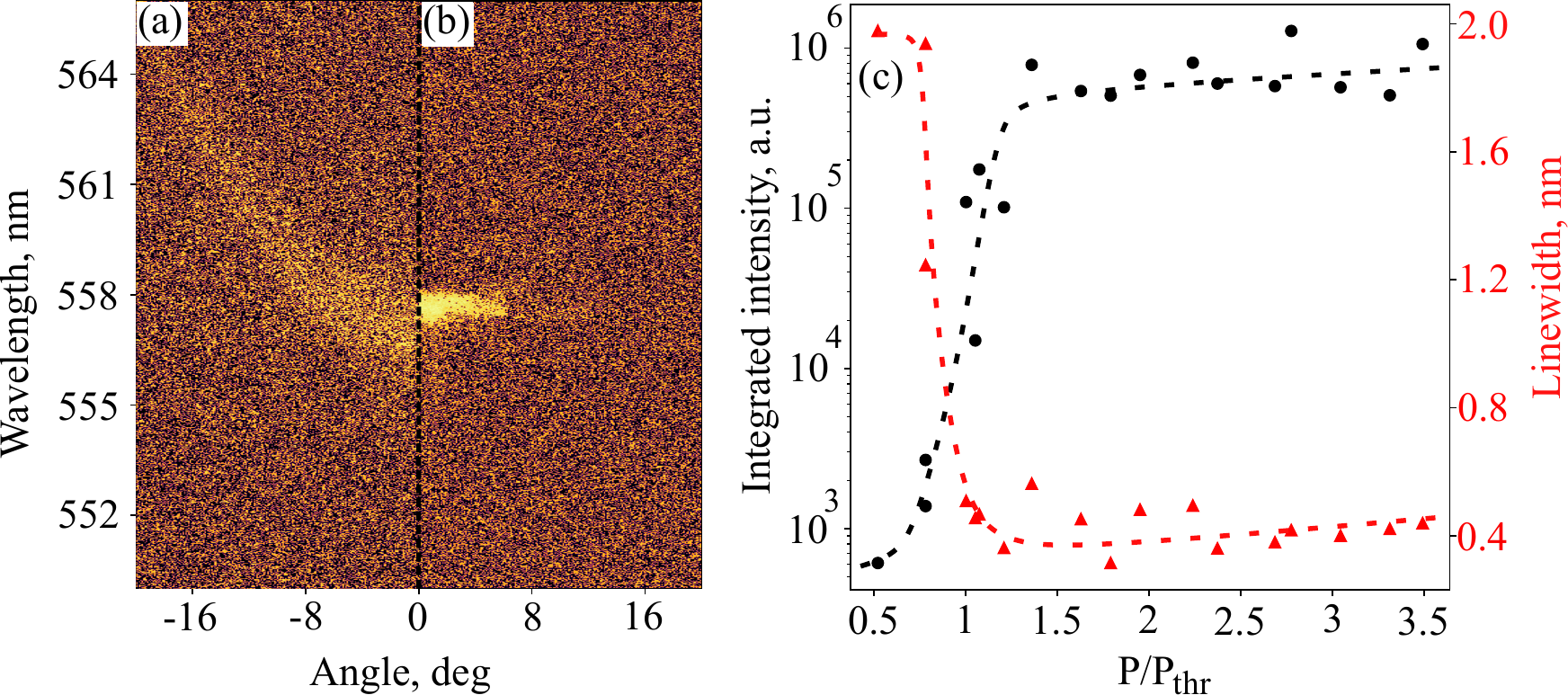}
  \caption{\label{fig:S5}
  \textbf{Threshold determination.} Angle-resolved (dispersion) images of polariton emission recorded \textbf{(a)} below threshold (accumulation over 10 pulses) and \textbf{(b)} above threshold (single shot). \textbf{(c)} Emission intensity integrated around normal incidence within $\pm10\degree$ (black circles) and condensate linewidth (red triangles) versus excitation density. The threshold is identified by the concomitant linewidth collapse and abrupt rise of intensity.}
\end{figure}

Figures~\ref{fig:S5}(a)$\&$(b) show angle-resolved (dispersion) images recorded below and above threshold, respectively. Figure~\ref{fig:S5}(c) presents the total emission intensity integrated around normal incidence within an angular window of $\pm10\degree$ (black circles) together with the emission linewidth (red triangles) versus excitation density. The threshold is identified by the concomitant linewidth collapse and abrupt rise of emission intensity. For a pump spot of $\sim23~\upmu\mathrm{m}$ (FWHM), we obtain the threshold pump fluence
\begin{equation}
  P_{\rm thr}=551~\upmu\mathrm{J\,cm^{-2}}.
\end{equation}

\section{Robustness of incompressible spectra extraction}
\label{sec:V}

This section documents analysis choices used to obtain the incompressible kinetic-energy spectrum $E_{\rm inc}(k)$ and provides basic robustness checks against common artifacts in finite-size, single-shot data.

\paragraph*{{\rm\textbf{Velocity field and masking.}}}
We compute the local wavevector field $\mathbf{k}(\mathbf{r})=\nabla\phi(\mathbf{r})$ from the reconstructed phase maps and restrict the analysis to pixels where the condensate intensity exceeds a fixed fraction of the shot maximum (default: 5\% as used in the main text). This mask suppresses noise-dominated regions and reduces edge artifacts.

\paragraph*{{\rm\textbf{Helmholtz decomposition.}}}
We obtain the incompressible component by performing a Helmholtz projection in $k$ space: for each Fourier component $\tilde{\mathbf{k}}(\mathbf{q})$ we subtract the longitudinal part parallel to $\mathbf{q}$, yielding $\tilde{\mathbf{k}}_{\rm inc}(\mathbf{q})=\tilde{\mathbf{k}}(\mathbf{q})-\mathbf{q}\,(\mathbf{q}\cdot\tilde{\mathbf{k}})/|\mathbf{q}|^2$, and then compute 
\begin{equation}
    E_{\rm inc}(q)\propto \sum_{\substack{|\mathbf{q}|\in [q,q+\Delta q]}} I(\mathbf{q})|\tilde{\mathbf{k}}_{\rm inc}(\mathbf{q})|^2,
\end{equation} summing over all $\mathbf{q}$, performing azimuthal averaging, and weighting each component by the local condensate intensity $I(\mathbf{q})$ \cite{koniakhin20202d}.

\paragraph*{{\rm\textbf{Robustness checks.}}}
We verified that the extracted local slope in the intermediate-$k$ region is stable under the following variations (data not shown):
(i) changing the intensity threshold from 5\% to 10\%;
(ii) modest Gaussian smoothing of $\phi(\mathbf{r})$ within the experimental resolution;
(iii) varying the $k$-binning used in the azimuthal average.
Within these variations, the fitted exponent $\beta$ in the $k^{-5/3}$-like segment remains within the quoted uncertainties in the main text, supporting that the observed scaling is not set by a single analysis parameter.


\end{document}